\documentstyle[12pt]{article}

\def\labelmark{}
\def\void{}

{\ifx\void\labelname\def\junk{\end{displaymath}}
\else\def\junk{\end{eqnarray}}\fi\junk\labelmark\def\labelname{}}

\newcommand{\bra}{\begin{array}}
\newcommand{\era}{\end{array}}
\newcommand{\beq}{\begin{equation}}
\newcommand{\eeq}{\end{equation}}
\newcommand{\bqn}{\begin{eqnarray}}
\newcommand{\eqn}{\end{eqnarray}}

\font\mybb=msbm10  at 12pt
\def\bb#1{\hbox{\mybb#1}}

\font\mybbi=msbm10  at 9pt
\def\bbi#1{\hbox{\mybbi#1}}

\def\Z{\bb Z}

\def\BC{\bb C}
\def\_\BC{\bbi C}
\def\RR{\bb R}
\def\Q{\bb Q}

\newcommand{\om}{\omega}
\newcommand{\Om}{\Omega}
\newcommand{\la}{\lambda}

\newcommand{\ep}{\epsilon}

\newcommand{\be}{\beta}
\newcommand{\ga}{\gamma}
\newcommand{\te}{\theta}

\newcommand{\pa}{\partial}
\newcommand{\al}{\alpha}
\newcommand{\ap}{\approx}
\newcommand{\de}{\delta}

\newcommand{\ka}{\kappa}

\newcommand{\ze}{\zeta}
\newcommand{\st}{\star}
\newcommand{\ti}{\tilde}
\newcommand{\da}{\dagger}

\newcommand{\ov}{\over}
\newcommand{\hb}{\hbar}
\newcommand{\vr}{\varphi}
\newcommand{\sq}{\sqrt}
\newcommand{\no}{\noindent}
\newcommand{\ev}{\equiv}
\newcommand{\lb}{\label}

\newcommand{\PL}[1]{ {\it Phys.~Lett.} {\bf #1}}

\begin{document}
\begin{titlepage}
\setcounter{page}{1}
\renewcommand{\thefootnote}{\fnsymbol{footnote}}

\begin{flushright}
Napoli DSF-NA- 15/2001\\
IC/2001/45\\
hep-th/0105303
\end{flushright}

\vspace{4mm}
\begin{center}

{\Large\bf Orbital Magnetism of Two-Dimension \\
Noncommutative Confined System} 
\vspace{8mm}

{\large\bf Ahmed Jellal
\footnote {On leave from {\it Feza G\"ursey Institute, P.O. Box 6, 81220 
\c{C}engelk\"oy, Istanbul, Turkey}}
\footnote {E-mail: {\textsf{jellal@gursey.gov.tr}}}}\\
\vspace{4mm}
{\em INFN-Sezione di Napoli, Dipartimento di Scienze Fisiche, Complesso Universitario\\ 
di Monte Sant'Angelo, Via Cintia ed. G, 80126 Naples, Italy}\\ 
{\textsf{jellal@na.infn.it}}\\
and\\
{\em High Energy Section, the Abdus Salam International Centre for Theoretical Physics\\
Strada Costiera 11, 34100, Trieste, Italy}\\
\vspace{4mm}
\textbf{Dedicated to Professor Erdal In\"on\"u on his 75 birthday}
\end{center}
\vspace{4mm}
\begin{abstract}
We study a system of spinless electrons moving in a two 
dimensional noncommutative space subject to a perpendicular 
magnetic field $\vec B$ and confined by a harmonic potential 
type ${1\over 2}mw_{0}r^2$. We look for the orbital magnetism 
of the electrons in different regimes of temperature $T$, magnetic 
field $\vec B$ and noncommutative parameter $\te$. We prove that 
the degeneracy of Landau levels can be lifted by the $\te$-term appearing 
in the electron energy spectrum at weak magnetic field. Using the 
{\it Berezin-Lieb} inequalities for thermodynamical potential, it is 
shown that in the high temperature limit, the system exibits a magnetic 
$\te$-dependent behaviour, which is missing in the commutative case. 
Moreover, a correction to susceptibility at low $T$ is observed. 
Using the {\it Fermi-Dirac} trace formulas, a generalization of the
thermodynamical potential, the average number of electrons and the 
magnetization is obtained. 
There is a critical point where the thermodynamical potential
becomes infinite in both of two methods above. So at
this point we deal with the partition function by adopting
another approach.  
The
standard results in the commutative case for this model can be
recovered by switching off the 
$\te$-parameter.
\end{abstract}
\end{titlepage}

\newpage

\section{Introduction}
 It seems that the noncommutativity appeared in physics since
{\it Palev} {\cite{pal1} investigated the noncanonical quantization of two
particles interacting via a harmonic potential {\it \`a la Wigner}
(see also {\cite{pal2,pal3,pal4,pal5}). One of the outcomes of his
approach is  
that the position of any one of the particles  cannot be localized 
in the space since  the coordinates of particles do not commute   
$$
[{\hat r}_i,{\hat r}_j]\neq 0\qquad i\neq j.
$$ 
In field theories, the noncommutativity is introduced by  
replacing the standard product by the star product. For a manifold 
parameterized by the coordinates $x^{i}$, the noncommutative relation can 
be written as follows {\cite{sei}}
\beq
[x^{i},x^{j}]=i\te^{ij},\label{(1)}
\eeq
where $\te^{ij}=\ep^{ij}\te$ is the noncommutative parameter
and is of dimension of $(length)^2$, 
$\ep^{12}=-\ep^{21}=1$. Basically, we are forced in this case
to replace $fg(x)=f(x)g(x)$ by the relation
\beq
f(x)*g(x)=\exp[{i\over 2}\te^{ij}\pa_{x^{i}}\pa_{y^{j}}]f(x)g(y)|_{x=y},
\label{2}
\eeq
where $f$ and $g$ are two arbitrary functions, supposed to be 
infinitely differentiable. The last equation defines the so-called 
the Moyal bracket of functions
\beq
\{f,g\}_{M.B.}=f*g-g*f,\label{3}
\eeq
which has been applied to solve some physical problems, 
for example see {\cite{zac}}.

Recently, some applications
of these mathematical tools were used to solve some physical problems. 
For instance, in quantum {\it Hall} effect a relation between 
$\te$ and the quantized {\it Hall} conductivity has been established 
{\cite{suss}} and a study of the multi-skyrmions near the filling
factor $\nu=1$ has been done \cite{lee}. 
Furthermore, in hydrogen atom spectrum the energy levels has been analyzed
in the framework of noncommutativity {\cite{jab}}. Subsequently, 
with {\it Dayi}  {\cite{jel1}}, we have considered the behaviour of
electrons in an external uniform magnetic field $\vec{B}$,
where the space coordinates perpendicular to $\vec{B}$ are taken 
as noncommuting. Calculating the susceptibility, we have found that  
the usual {\it Landau} diamagnetism is modified. We have also computed the susceptibility 
according to nonextensive statistics. 
We have found that these two methods agree under certain conditions. Basically,
this paper {\cite{jel1}} can offer some possibilities to give an
noncommutative description for any system showing an anomaly 
in~the~{\it~Boltzmann-Gibbs}~theory~related~to~statistical~physics.\\

On the other hand, orbital magnetism, which is possible only in
quantum mechanics, has stimulated some work in this 
period (see \cite{ifu} and references therein). 
With {\it Gazeau et al} {\cite{jel2}}, we have studied  
the possible occurrence of orbital magnetism 
for two-dimensional electrons confined by 
a harmonic potential in various regimes
of temperature and magnetic field. Standard coherent state 
families are used for 
calculating symbols of various involved observables like 
the thermodynamical potential, 
magnetic moment, or the spatial distribution of the current. 
Their expressions are given in a closed form and 
the resulting {\it Berezin-Lieb} 
inequalities provide a straightforward way to 
study magnetism in various limit
regimes. In particular, we have predicted a 
paramagnetic behaviour in the thermodynamical limit
as well as in the quasiclassical limit under a weak field. 
Finally, we have obtained an exact expression for the magnetic moment 
which yields
a full description of the phase diagram of the magnetization. \\ 

Our main goal in this paper is to study the orbital magnetism of the
model used in \cite{jel2} in noncommutative space.  
Our idea is to consider a system of electrons moving on a
noncommutative space and subject to a perpendicular magnetic field
and to harmonic confining potential. We show the 
differences of the commutative and noncommutative
cases. In particular, employing
the {\it Berezin-Lieb} inequalities 
we find that there is no
degeneracy when the magnetic field is weak and 
point out a correction to susceptibility 
at low and high temperature~$T$.
Furthermore, using 
the {\it Fermi-Dirac} trace formulas, a general expression
is derived for the thermodynamical potential, 
the average number of
electrons and the magnetization. A critical
point is found, such that at ${eB\te\ov c}=-2$, the
thermodynamical potential becomes infinite in both of
two methods mentioned above. However,
by using another approach, we obtain the 
thermodynamical potential, which is found
to be equivalent to that of
$2d$ electrons in a uniform magnetic field. As a
consequence we find
infinite susceptibility for zero magnetic field.  

The outline of the paper is as follows. In section $2$, we give the 
noncommutative
version of a Hamiltonian describing $2d$-electrons in 
the presence of a
perpendicular magnetic
field and confining potential. Using two different methods, we investigate
the energy spectrum
and the corresponding eigenfunctions in
section $3$. We study the degeneracy of {\it Landau} levels 
in section~$4$, where we also start with
the realization of  some algebras and  
investigate the magnetic field limits. 
In section $5$, we derive the thermodynamical 
potential and the related physics quantities by using 
two methods: the first makes use of the {\it Berezin-Lieb} inequalities
and the second one employs the 
{\it Fermi-Dirac} trace formulas. 
At critical point, we use another approach 
to obtain the thermodynamical potential and related quantities. 
The final section
is devoted to conclusions and perspectives. 

\section{Electron in noncommutative space}
 Let us consider a system of spinless 
electrons $(m,e)$ living on the $(x,y)$-space in a
magnetic field ${\vec B}$. We recall that the eigenstates and 
eigenvalues were investigated for the first 
time by {\it Landau} {\cite{lan}}. 
When a harmonic confining potential is introduced 
and the {\it Coulomb} interactions are neglected, this system is 
described by the {\it Fock-Darwin} Hamiltonian \cite{foc,dar,jac}
\beq
H={1\over 2m}\Big({\vec P}+{e\over c}{\vec A}\Big)^2+{1\over 2}mw_{0}^2r^2,
\label{h1}
\eeq
where ${\vec P}$ is the canonical momentum and ${\vec A}$ is
the vector potential. We will study this Hamiltonian
by making use of the commutation relations
\beq
[x^i,p^j]=i\hb\de^{ij},\qquad [p^i,p^j]=0,\lb{ps}
\eeq
as well as eq.(1), and by choosing the symmetric gauge
\beq
{\vec A}=\Big(-{B\over 2}y,{B\over 2}x,0\Big).\label{sg}
\eeq
According to this recipe, the above Hamiltonian acts
on an arbitrary function $\Psi(\vec{r},t)$ as
\beq
\bra{l}
H \st \Psi (\vec{r},t) =\frac{1}{2m}\left[
\left( p_x -\frac{eB}{2c} y \right)^2 +
\left( p_y +\frac{eB}{2c} x \right)^2 
+m^2w_{0}^2(x^2+y^2) \right]\st\Psi (\vec{r},t)\\
\qquad\qquad\;\;\;\; \equiv H_{\te} \Psi (\vec{r},t).\label{h2}
\era
\eeq
Therefore, the noncommutative version of eq.(\ref{h1}) 
can be written as follows
\beq
H_{\te}={1\over 2m}\Bigg(\Big({\ti p}_x-{eB\ov 2c}y\Big)^2+
\Big({\ti p}_y+{eB\ov 2c}x\Big)^2\Bigg)+
{1\over 2}mw_{0}^2\Bigg(\Big({1 \ov 2}\te p_x+y\Big)^2+
\Big({1\over 2}\te p_y-x\Big)^2\Bigg).\label{nch}
\eeq
Here ${\ti p}_{\mu}$ is a linear function of 
the noncommutative parameter, such that
\beq
{\ti p}_{\mu}=(1+{m\om_c\ov 4}\te)p_{\mu}, \qquad \mu=x,y.\label{ncm}
\eeq
This problem has been analyzed without 
the confining potential and at 
noncommutative level   
on the torus {\cite{pol1}}. Notice that when $\te$ vanishes, the standard 
Hamiltonian can be recovered.

To close this section, we mention that the Hamiltonian eq.(\ref{h1})
has been considered on the noncommutative space \cite{pol2}, 
where the relation
\beq
[p_x,p_y]=iB,\label{pncm}
\eeq
and the convention $(e=1,c=1)$ are used, which is 
not the case for our analysis. However, we can find 
identical results concerning the noncommutative Hamiltonian
formalism if we make a redefinition of the magnetic field,
such that
\beq
B_{P}=-B_{J}(1+{\te\ov 4}B_{J}),\label{pj}
\eeq
where $B_{P}$ is the magnetic field used by {\it Polychronakos et al}
and $B_{J}$ is the one appearing in our formulas
\footnote{I'm grateful to A. Polychronakos for pointing
eq.(\ref{pj}) out on May 31, 2001}.

\section{Eigenstates and eigenvalues of $H_{\te}$} 
We adopt two methods to obtain the energy spectrum  and the eigenstates 
of $H_{\te}$. The first one utilizes {\it Weyl-Heisenberg} symmetries
and the last one is related to the stationary {\it Schr\"odinger} 
equation.
 
\subsection{Algebraic method}
It is possible to write the noncommutative Hamiltonian as the sum of
two independent harmonic 
oscillator Hamiltonians ${\ti H}_{0}$ plus the angular 
momentum operator on $z$-direction
$L_z$. Therefore, we have
\beq
H_{\te}= {\ti H}_{0}+{{\ti\om}_{c}\ov 2}L_z,
\label{9}
\eeq
where ${\ti H}_{0}$ and $L_z$ are given by
\beq
\bra{l}
{\ti H}_{0}={1\ov 2m}\Big({\hat p}_{x}^2+
\frac{1}{8} m \om^2 x^2\Big)+{1\ov 2m}\Big({\hat p}_{y}^2+\frac{1}{8} 
m \om^2 y^2\Big),\\
L_z = xp_y - yp_x .\label{10}
\era
\eeq
Here $\om_c = eB/mc$ is the cyclotron frequency, 
$\om = \sqrt{\om_c^2+ 4\om_0^2}$ and 
\beq
{\hat p}_{\mu}^2=(1+{m\om_c\ov 2}\te+({m\om\ov 4}\te)^2)p_{\mu}^2, 
\qquad {\ti\om}_{c}=
\om_{c}(1+({\om_c\over 4}-{\om_0^2\ov\om_c})m\te).
\label{11}
\eeq
We want to express $H_{\te}$ in terms of creation and annihilation 
operators. For that, we introduce the following operators in the complex 
plane $(z,{\bar z})$
\beq
\bra{l}
{\ti a}_{d}=\frac{1}{{2}}({\ti\xi} {\bar z}+
\frac{i}{2\hbar{\ti\xi}}p_z),
\qquad {\ti a}_{d}^{\da}=\frac{1}{{2}}({\ti\xi} z-
\frac{i}{2\hbar{\ti\xi}}p_{\bar z}),\\
{\ti a}_{g}=\frac{1}{{2}}({\ti\xi} z+ 
\frac{i}{2\hbar{\ti\xi}}p_{\bar z}),
\qquad {\ti a}_{g}^{\da}=\frac{1}{{2}}({\ti\xi} {\bar z}-
\frac{i}{2\hbar{\ti\xi}}p_z),
\label{12}
\era
\eeq
where ${\ti\xi}$ is a $\te$-function, such that
\beq
{\ti\xi} = \root 4 \of{({m\om/2\hb})^2\ov 1+{m\om_c\ov 2}\te+
({m\om\ov 4}\te)^2}.\label{13}
\eeq
It is easy to show that 
\beq
[{\ti a}_{d},{\ti a}_{d}^{\da}]=1=[{\ti a}_{g},{\ti a}_{g}^{\da}],
\label{14}
\eeq
and other commutators vanish. Consequently,  ${\ti H}_{0}$ and $L_z$ take the 
new forms 
\beq 
{\ti H}_{0} = {\hb{\ti\om}\ov 2}\; 
({\ti N}_{d}+{\ti N}_{g}+1), \ \ L_z = \hb ({\ti N}_{d} -{\ti N}_{g}),
\label{18}
\eeq
where ${\ti N}_{d} ={\ti a}_{d}^{\da}{\ti a}_{d}$, 
${\ti N}_{g}= {\ti a}_{g}^{\da}{\ti a}_{g}$
are the number operators and ${\ti\om}$ is $\te$-dependent:
\beq
{\ti\om}=\om\sq{1+{m\om_c\ov 2}\te+({m\om\ov4}\te)^2}\label{19}.
\eeq
Actually, we have the following expression 
for the noncommutative Hamiltonian 
\beq 
H_{\te}={\hb{\ti\om}\ov 2}\; 
({\ti N}_{d}+{\ti N}_{g}+1)+{\hb{\ti\om}_{c}\ov 2}\;({\ti N}_{d}- 
{\ti N}_{g}).
\label{20}
\eeq
This latter can be arranged as follows 
\beq 
H_{\te}={\hb\ov 2}\; 
({\ti N}_{d}{\ti\om}_{+}+{\ti N}_{g}{\ti\om}_{-}+{\ti\om}),
\label{21}
\eeq
and we have 
\beq
\bra{l}
{\ti\om}_{\pm}=\om\sq{1+{m\om_c\ov 2}\te+({m\om\ov4}\te)^2}\pm
\om_{c}(1+({\om_c\over 4}-{\om_0^2\ov\om_c})m\te)\\
\;\;\;\;\; ={\ti\om}\pm{\ti\om}_{c}.\label{22}
\era
\eeq
We derive immediately the energy spectrum from the relation
\beq 
{\ti H}_{\te}\mid {\ti n}_{d},{\ti n}_{g}\rangle=E_{{\ti n}_{d}{\ti n}_{g}}
\mid {\ti n}_{d},{\ti n}_{g}\rangle,\label{23} 
\eeq
which leads to
\beq
\bra{l}
E_{{\ti n}_{d}{\ti n}_{g}}={\hb\ov 2}({\ti n}_{d}{\ti\om}_{+}+{\ti n}_{g}
{\ti\om}_{-}+{\ti\om}).
\label{24}
\era
\eeq
${\ti n}_{d}$ and ${\ti n}_{g}$ are non-negative integers. The corresponding 
eigenstates are tensor products of single {\it Fock} oscillator states:
\beq 
\mid {\ti n}_{d},{\ti n}_{g}\rangle=\frac{1}{\sqrt{{\ti n}_{d}!{\ti n}_{g}!}}
({\ti a}_{d}^{\dagger})^{{\ti n}_{d}}({\ti a}_{g}^{\dagger})^{{\ti n}_{g}} 
\mid {\ti 0}, {\ti 0} \rangle.\label{25}
\eeq
$\mid {\ti 0}, {\ti 0} \rangle$ is the vacuum of $H_{\te}$. Noting that
if we use eq.(\ref{pj}), we recover the results obtained in
\cite{pol2} for the Hamiltonian eq.(\ref{h1}) in noncommutative
space.
 
\subsection{Analytical method}
To obtain the analytical solutions of the present problem, we introduce 
the polar coordinates $(x,y)=(r\sin\vr,r\cos\vr)$, with $0<r<\infty$
and $0\le\vr\le \pi$. In this case, 
the stationary {\it Schr\"{o}dinger} 
equation can be written as follows
\beq
\bra{l}
\Bigg(-{\hb^2\ov 2m}(1+{m\om_c\ov 2}\te+
({m\om\ov 4}\te)^2)\Big(\pa_r^2+\frac{1}{r}\pa_r +\frac{1}{r^2}
\pa_{\vr}^2\Big)-i{\hb{\ti\om}_c\ov 2}
\pa_{\vr}+\frac{m}{8}\om^2 r^2\Bigg)\Psi_{\te}(r,\vr)= E_{\te}\Psi_{\te}(r,\vr).\label{26}
\era
\eeq
Notice that $H_{\te}$ and $L_z$ commute. Therefore, following the 
fundamental
principle of quantum mechanics, these operators have a common basis
of eigenvectors. 
Then, by choosing these eigenfunctions as $\Psi_{\te}(r,\vr)= 
R_{\te}(r)e^{i\al\vr}$, 
we can show that eq.(26) yields
\beq
\Big(\pa_r^2+\frac{1}{r}\pa_r -\frac{\al^2}{r^2}\Big)R_{\te}(r)
-\Big({\ti k}^2-{\ti\ze}_2r^2\Big)R_{\te}(r)=0, \label{27}
\eeq
where 
\beq 
\bra{l}
{\ti k}^2= {E_{\te}-{\hb\om_c\ov 2}\Big(1+({\om_c\over 4}-
{\om_0^2\ov\om_c})m\te\Big)\al
\ov{\hb^2\ov 2}\Big(1+{m\om_c\ov 2}\te+({m\om\ov 4}\te)^2\Big)}, \qquad
{\ti\ze}^2={({m\om/2\hb})^2\ov 1+{m\om_c\ov 2}\te+({m\om\ov 4}\te)^2}.\lb{28}
\era
\eeq
By straighforward computation we show that 
\beq
R_{\te}(r)=r^{|\al|}\exp\Big(-{{\ti\ze} r^2\ov 2}\Big)L_{\te}(r)
\lb{29}
\eeq
is a solution of the above equation, where the $L_{\te}(r)$ are the 
{\it Laguerre} polynomials obeying
\beq
\pa_r^2L_{\te}+\Big({2|\al|+1\ov r}-2\al r\Big)\pa_rL_{\te}-
\Big(2{\ti\ze}(|\al|+1)-{\ti k}^2\Big)L_{\te}=0.\lb{30}
\eeq
Therefore, we can obtain the explicit eigenstates of $H_{\te}$ as
\beq
\Psi_{\te}(r,\vr)=\Psi_{n,\al,\te}(r,\vr)
=(-1)^n\;{\sqrt{{\ti\ze}\ov\pi}} \;\sqrt{\frac{n!}{(n+\vert\al\vert)!}} 
\exp(-{{\ti\ze} r^2\ov 2})\; \Big({\sqrt{{\ti\ze}}r}\Big)^{\vert 
\al \vert}\, L_{n,\te}^{(|\al|)}\Big({{\ti\ze} r^2}\Big)\;e^{i\al\vr},
\lb{31} 
\eeq
where 
$ L_{n,\te}^{(|\al|)}\Big({{\ti\ze} r^2}\Big)=\sum_{m=0}^{n}(-1)^m 
\pmatrix{n+|\al|\cr n-m\cr}{\Big({{\ti\ze} r^2}\Big)^m\ov m!}$,
$n=0, 1, 2,\cdots$ is the principal quantum number and
$\al=0, \pm 1, \pm 2 , \cdots$ is the angular momentum quantum number.
The corresponding spectrum is given by
\beq
E_{n\al,\te} ={\hb{\ti\om}\ov 2}\;(n+\frac{\vert \al \vert+1}{2})+
{\hb{\ti\om}_{c}\ov 2}\al.\lb{32}
\eeq
Since $L_{n,\te}^{(|\al|)}(0)={(n+\al)(n+\al-1)...(\al+1)\ov
n!}$, then from eq.(31) we observe immediately that 
$\Psi_{n,\al,\te}(0)=~0$ and also becomes zero when $r$
goes to infinity.
Returning now to the algebraic method, 
we can see that $n$ and $\al$ are connected to 
${\ti n}_{d}$ and ${\ti n}_{g}$ by  
$$
{\ti n}_{d}=n+ \frac{1}{2}(\vert \al \vert + \al), \ \mbox{and}\
{\ti n}_{g}=n+ \frac{1}{2}(\vert \al \vert - \al).
$$
Notice that
$\Psi_{n,\al,\te}(r,\vr) =\langle r , \vr \mid n, \al \rangle 
               =\langle r , \vr \mid {\ti n}_{d}, {\ti n}_{g} \rangle$.

\section{Degeneracy of Landau levels}
As in the commutative case \cite{jel2}, we can give a realization of certain algebras, 
in particular $su(2)$ and $su(1,1)$, in terms of the creation and
annihilation operators 
defined before. We can also study some particular cases in which
the magnetic field takes some limiting values.

\subsection{Algebras $su(2)$ and $su(1,1)$}
We start with the former one. The algebra generators can be built as
\beq
{\ti S}_+ = {\ti a}_{d}^{\dagger} {\ti a}_{g}, \ \  
{\ti S}_-= {\ti a}_{g}^{\dagger} {\ti a}_{d}, \ \  
{\ti S}_{z} = \frac{{\ti N}_{d}-{\ti N}_{g}}{2}=\frac{L_z}{2\hbar}.\lb{33}
\eeq
It easy to show that these generators verify the following 
commutation relations
\beq
\lbrack {\ti S}_+ ,{\ti S}_-\rbrack = 2{\ti S}_{z}, \ \  
\lbrack {\ti S}_{z},{\ti S}_{\pm} \rbrack = \pm {\ti S}_{\pm}. \lb{34}
\eeq
Subsequently, we can define also the invariant Casimir operator 
in terms of $su(2)$ generators 
\beq
{\ti{\cal C}} = \frac{1}{2}({\ti S}_+ {\ti S}_- + {\ti S}_- {\ti S}_+) 
+ {\ti S}_{z}^2 
= (\frac{{\ti N}_{d} + {\ti N}_{g}}{2})(\frac{{\ti N}_{d} + 
{\ti N}_{g}}{2} + 1).
\lb{35} 
\eeq
We prove that $H_{\te}$ is not invariant under this algebra.  
As in the commutative case, for a given value 
$\ga=({\ti n}_{d}+{\ti n}_{g})/2$, there exists a 
$(2\ga + 1)$-dimensional UIR of $su(2)$ in which the
operator $ {\ti S}_{z}$ has its spectral values in the range 
$ -\ga\leq \rho= ({\ti n}_{d} - {\ti n}_{g})/2 \leq \ga$.

Following the same idea the other algebra can be realized as follows
\beq
{\ti T}_+ = {\ti a}_{d}^{\dagger} {\ti a}_{g}^{\dagger}, \ \ 
{\ti T}_- = {\ti a}_{d} {\ti a}_{g}, \ \
{\ti T}_0 = \frac{1}{2}({\ti N}_{d} + {\ti N}_{g}+ 1)= 
{{\ti H}_{0}\ov\hbar{\ti\om}}.\lb{36}
\eeq
Then we reproduce the commutation relations generating 
the $su(1,1)$ algebra:
\beq
\lbrack {\ti T}_{+},{\ti T}_{-} \rbrack = -2{\ti T}_{0}, \ \
\lbrack {\ti T}_{0},{\ti T}_{\pm} \rbrack = \pm {\ti T}_{\pm}.
\lb{37}
\eeq
Furthermore, its  Casimir operator is given by
\beq
{\ti{\cal D}}=\frac{1}{2}({\ti T}_+{\ti T}_- + {\ti T}_-{\ti T}_+) - 
{\ti T}_0^2 
        =-(\frac{{\ti N}_{d}-{\ti N}_{g}}{2}+\frac{1}{2})
(\frac{{\ti N}_{d}-{\ti N}_{g}}{2}-\frac{1}{2})
        =-\frac{1}{4}(\frac{L_z}{\hbar}^2 -1). \lb{38}
\eeq
This algebra also is not a symmetry of the noncommutative 
Hamiltonian.
Notice that, when ${\ti n}_{d} \geq {\ti n}_{g}$,  
for a given value $\eta = ({\ti n}_{d}-{\ti n}_{d}+ 1)/2 \geq 1/2 $, 
there exists a UIR of $su(1,1)$ in the discrete series, 
in which the operator $ {\ti T}_{0}$ has its spectral 
values in the infinite 
range $ \eta, \eta +1, \eta + 2, \cdots$. However,
when ${\ti n}_{d} \leq {\ti n}_{g}$, for a given value 
$\vartheta = (-{\ti n}_{d}+{\ti n}_{g}+1)/2 \geq 1/2$, 
there also exists a UIR of $su(1,1)$ in which the spectral value of the
operator $ {\ti T}_{0} $ runs in 
the~infinite~range~$~\vartheta,~\vartheta~+1,~\vartheta~+~2,~\cdots$.\\

\subsection{Magnetic field limits}
Let us examine some particalur cases of 
the magnetic field: weak field and strong field limits.
We begin by arranging the energy spectrum as follows
\beq
{\ti E}_{n\al} ={\hb{\ti\om}\ov 2}\;\ga+
{\hb{\ti\om}_{c}\ov 2}\;\rho+{\hb{\ti\om}\ov 2}.\lb{41}
\eeq

\no{\it{i-{\underline{Weak field case}}}}\\
Suppose that $\om_c \ll \om_0$, then the above equation can be 
approximated by
\beq
{\ti E}_{n\al}\ap\hb\om_0\sq{1+({m\om_0\ov 2}\te)^2}\;
(2\ga+ 1)-{\hb m\om_0^2\ov 2}\te\;\rho \ev E_{\ga,\rho},\lb{39}
\eeq
which tells us there is no degeneracy of 
{\it Landau} levels. This effect
is due to the presence of the $\te$-term in 
the energy spectrum eq.(40). This 
latter shows a difference with the commutative case, where we have 
pointed out {\cite{jel2}} that $su(2)$ is behind the degeneracy of 
{\it Landau} levels at weak field.\\

\no{\it{ii-{\underline{Strong field case}}}}\\
In the limit of strong magnetic field $\om_c \gg \om_0$,
we have 
\beq
E_{{\ti n}_{d}{\ti n}_{g}} \approx \hbar 
\om_c(1+{m\om_c\ov 4}\te)({\ti n}_{d}+ \frac{1}{2}).
\lb{40} 
\eeq
As in the commutative case by redefining $\om_c$ we get harmonic 
oscillator and it is still true that 
for a given value of ${\ti n}_{d}$, we have an infinite 
degeneracy labelled by ${\ti n}_{g}$ or 
by $\al = {\ti n}_{d} - {\ti n}_{g} \leq {\ti n}_{d}$. 
The quantum number ${\ti n}_{d}$ corresponds to the 
{\it Landau} level index
(as well as $n$ for negative $\al$).  
One can reinterpret it in terms of $su(1,1)$ symmetry by noting that, 
for a given value of $\al \leq 0$,  
the energy eigenstates are ladder states for the discrete series 
representation labelled by $\vartheta = - \al/2 + 1/2$.\\

\no{\it{iii-{\underline{Generic intermediate case}}}}\\
We distinguish two cases: For ${\ti\om}_+/{\ti\om}_- 
\notin \Q$, what we can do is just to write
the energy spectrum in the form  
\beq
{\cal E}_{{\ti n}_d {\ti n}_g} \equiv 
\frac{E_{{\ti n}_d {\ti n}_g}}{\hbar{\ti\om}_-}-\frac{{\ti\om}}
{2{\ti\om}_-}= \frac{{\ti\om}_+}{{\ti\om}_-}{\ti n}_d + {\ti n}_g,
\eeq
otherwise there is no information about degeneracy. 
For ${\ti\om}_+/{\ti\om}_- = p/q \in \Q$, this latter is possible:
\beq
E_{{\ti n}_d {\ti n}_g} = E_{{\ti n}'_d {\ti n}'_g} \ \mbox{iff} \ 
\frac{p}{q} = - \frac{{\ti n}_g - {\ti n}'_g}{{\ti n}_d - {\ti n}'_d}.
\eeq 

\no{\it{iv-{\underline{Conclusion}}}}\\
The above analysis leads us to conclude that 
the introduction of the
noncommutative parameter can solve 
some problems. For instance the
degeneracy of the {\it Landau} levels 
is lifted via the $\te$-term for a weak
magnetic field. Indeed, for any non-zero $\te$-value,
the term ${\hb m\om_0^2\ov 2}\te\;\rho$ 
is present in eq.(40), which means 
that for any given eigenvalue 
$E_{\la,\rho}$ there is only one 
eigenfunction parametrized by 
the same integers $\la$ and $\rho$. 
However for $\te=0$, we recover 
the harmonic oscillator
for $2-$dimensional {\it Landau} problem, 
which is a degenerate system. 
 
\section{Thermodynamical potential}
We make the assumption that the total 
number $\langle {\ti N}_e \rangle$ of 
electrons is large enough so that 
the difference between a grand canonical
ensemble and a canonical one is not of 
importance \cite{ifu,jel2}. Then, the
thermodynamical 
potential can be written as follows 
\beq
\Om_{\te} = -\frac{1}{\be}\mbox{Tr}
\log{(1+e^{-\be(H_{\te}-\mu)})} ,\label{55}
\eeq
with $\be = 1/(k_BT)$. Evaluating 
the trace on the derived eigenstates, we obtain
\beq
\Om_{\te}=\sum_{{\ti n}_{d}, {\ti n}_{g}}^{\infty} 
\log{(1+e^{-\be({\hbar\ov 2}({\ti\om}_+
{\ti n}_{d}+{\ti\om}_-{\ti n}_{g}+{\ti\om})-\mu)})}.
\eeq
By definition, the magnetic moment $M_{\te}$ is 
\beq
M_{\te} = -\left(\frac{\pa \Om_{\te}}{\pa B}\right)_{\mu}, \label{54}
\eeq
and the average number of electrons is given by
\beq
\langle {\ti N}_e\rangle= -\partial_{\mu} \Om_{\te}.\lb{56}
\eeq

On the other hand, it is not easy to manipulate directly eq.(45) and 
subsequently
eqs.(46-47). Basically, we need some tools to do that; this is the
reason why we introduce coherent states \cite{kla,gaz}. Then, 
before investigating the thermodynamical potential, we start with
constructing 
the coherent states. Note that this construction is, more
or less, the same
as in the standard case.
 
\subsection{Coherent States}
Using standard methods, the coherent states for the present system can
be constructed as follows
\beq
\mid {\ti z}_{d},{\ti z}_{g}\rangle = \exp{\lbrack-\frac{1}{2}
(\vert {\ti z}_{d} \vert^2+\vert {\ti z}_{g}\vert^2)\rbrack }\,
 e^{{\ti z}_{d} {\ti a}_{d}^{\dagger} + 
{\ti z}_{g} {\ti a}_{g}^{\dagger}} 
\mid {\ti 0}, {\ti 0}\rangle .\lb{42}
\eeq 
It easy to observe that 
\beq
{\ti a}_{d} \mid {\ti z}_{d}, {\ti z}_{g} \rangle = {\ti z}_{d} 
\mid {\ti z}_{d}, {\ti z}_{g}\rangle,\ \ {\ti a}_{g} \mid {\ti z}_{d}, 
{\ti z}_{g} \rangle = {\ti z}_{g}\mid {\ti z}_{d}, {\ti z}_{g} \rangle.
\eeq
We cite some interesting properties, which will be useful in the next.
The first one is the action identity:
\beq
\check{H_{\te}}({\ti z}_{d}, {\ti z}_{g})
\equiv\langle {\ti z}_{d}, {\ti z}_{g}\mid H_{\te}\mid {\ti z}_{d}, 
{\ti z}_{g}\rangle ={\hbar\ov 2}\Big({\ti\om}_{+}\vert {\ti z}_{d}\vert^2+
{\ti\om}_{-}\vert {\ti z}_{g}\vert^2+{\ti\om}\Big).\lb{43}
\eeq
In the literature, the function $\check{ H_{\te}}({\ti z}_{d}, {\ti z}_{g})$ is 
known as the lower (resp. contravariant) symbol of the operator 
$H_{\te}$ \cite{ber,fks}. 
It will play an important role in the present context.
The second one is the resolution of the unity:
\beq
\mbox{I}=\frac{1}{\pi^2}\int_{\BC^2}\mid {\ti z}_{d}, {\ti z}_{g}
\rangle\langle {\ti z}_{d}, {\ti z}_{g} \mid\,
          d^2{\ti z}_{d} \, d^2{\ti z}_{g}.
\eeq
The last property is also crucial in our context. Indeed, 
For any observable $A$ with suitable operator properties (trace-class,
...), 
there exists a unique upper (or covariant) symbol $\hat{A}({\ti z}_{d}, 
{\ti z}_{g})$ defined by
\beq
A=\frac{1}{\pi^2}\int_{\BC^2}\, \hat{A}({\ti z}_{d}, {\ti z}_{g})\, 
\mid {\ti z}_{d}, {\ti z}_{g}\rangle\langle {\ti z}_{d}, {\ti z}_{g} \mid\, 
d^2{\ti z}_{d} \, d^2{\ti z}_{g}.\lb{44}
\eeq
It is easy to see that the upper symbols for number operators are  
\beq
{\hat{\ti N}}_{d}({\ti z}_{d}, {\ti z}_{g}) = \vert {\ti z}_{d}\vert^2 -1,\ \ 
{\hat{\ti N}}_{g}({\ti z}_{d}, {\ti z}_{g}) = \vert  {\ti z}_{d}\vert^2 -1,
\lb{45} 
\eeq
which imply the following one for the noncommutative Hamiltonian 
\beq
\hat{H_{\te}}({\ti z}_{d},{\ti z}_{g})={\hbar\ov 2}\Big({\ti\om}_{+}
\vert {\ti z}_{d}\vert^2+{\ti\om}_{-}\vert {\ti z}_{g}\vert^2-
{\ti\om}\Big). \lb{46}
\eeq
Notice that there is another useful trace identity for a trace-class
observable $A$, such that
\beq
\mbox{Tr}A =\frac{1}{\pi^2}\int_{\BC^2}\, \check{A}({\ti z}_{d},{\ti z}_{g})\, 
d^2{\ti z}_{d} \, d^2{\ti z}_{g}
=\frac{1}{\pi^2}\int_{\BC^2}\, \hat{A}({\ti z}_{d},{\ti z}_{g})\, 
d^2{\ti z}_{d} \, d^2{\ti z}_{g},\lb{47}
\eeq 
where we have $\check{A}({\ti z}_{d},
{\ti z}_{g})\equiv\langle {\ti z}_{d},{\ti z}_{g}
\mid  A\mid {\ti z}_{d},{\ti z}_{g}\rangle$.

\subsection{Berezin-Lieb inequalities}
Let us observe that $\log{(1 + e^{-\be({H}_{\te} - \mu)})}$ 
is a convex function 
of the positive Hamiltonian ${H}_{\te}$. Then, the {\it Berezin-Lieb}  
inequalities can be applied to study the quasi-classical behaviour of 
the thermodynamical potential. For any convex function $g(A)$ of the
observable 
$A$ it is possible to write  \cite{ber,fks} 
\beq
\frac{1}{\pi^2}\int_{\BC^2}g(\check{A})\,d^2{\ti z}_{d}\,d^2{\ti z}_{g}
\leq \mbox{Tr}g(A) \leq 
\frac{1}{\pi^2}\int_{\BC^2}g(\hat{A}) \, d^2{\ti z}_{d}\,d^2{\ti z}_{g}.
\label{58}
\eeq
This formula can be used for evaluating the (concave) thermodynamical
potential. Then,
we have
\beq
-\frac{1}{\be\pi^2}\int_{\BC^2}\log{(1+e^{-\be(\hat{{H}}_{\te}
-\mu)})}\,d^2{\ti z}_{d}\,d^2{\ti z}_{g} \leq \Om_{\te} \leq  
-\frac{1}{\be\pi^2}\int_{\BC^2}\log{(1+e^{-\be(\check{{H}}_{\te}-\mu)})}\, 
d^2{\ti z}_{d}\, d^2{\ti z}_{g}.\label{59}
\eeq
Using eqs.(50) and (54) and performing the angular 
integrations, we get
\bqn
\nonumber 
-\frac{1}{\be}\int_{0}^{\infty}d{\ti u}_{d}\,\int_{0}^{\infty} d{\ti u}_{g}\,
\log{(1+e^{-\be({\hbar\ov 2}({\ti\om}_+{\ti u}_{d}+{\ti\om}_-{\ti u}_{g}-
{\ti\om})-\mu)})}\leq\Om_{\te},\\
\Om_{\te} \leq -\frac{1}{\be}
\int_{0}^{\infty}d{\ti u}_{d}\,\int_{0}^{\infty}d{\ti u}_{g}\,
\log{(1+e^{-\be({\hbar\ov 2}({\ti\om}_+{\ti u}_{d}+{\ti\om}_-{\ti u}_{g}+
{\ti\om})-\mu)})}, \label{60}
\eqn
where $ {\ti u}_{d}= \vert {\ti z}_{d} \vert^2$ and ${\ti u}_{g} = 
\vert {\ti z}_{g} \vert^2$. In order to calculate the last integrals, we put 
${\ti u}~=~{\be\hbar\ov 2}({\ti \om}_+{\ti u}_{d}+{\ti \om}_-{\ti 
u}_{g},\, 
{\ti v}={\be\hbar\ov 2}{\ti\om}_+{\ti u}_{d}$, then
performing an integration by parts, and introducing the control parameters 
${\ti\ka}_{\pm} =\exp{(\be(\mu \pm \hbar {\ti\om}/2))}$, we obtain 
\beq
{\phi}({\ti\ka}_+)\leq\Om_{\te}\leq{\phi}({\ti\ka}_-), \label{61}
\eeq
where ${\phi}({\ti\ka})$ takes the form
\bqn
\nonumber 
{\phi}({\ti\ka})&=& -\frac{2{\ti\ka}}
{\be(\be\hbar)^2{\ti\om}_+{\ti\om}_-}
\int_0^{\infty}\frac{{\ti u}^2 e^{-{\ti u}}}
{1+{\ti\ka} e^{-{\ti u}}}\, d{\ti u} \\
&=& \left\{ \begin{array}{ll} 
 \frac{4}{\be(\be\hbar)^2{\ti\om}_+{\ti\om}_-}\,
{\ti F}_{3}(-{\ti\ka}) & \mbox{for ${\ti\ka} \leq 1$},\\
 \frac{4}{\be(\be\hbar)^2{\ti\om}_+{\ti\om}_-}\left\lbrack 
-\frac{(\log{{\ti\ka}})^3}{6}
-\frac{\pi^2\log{{\ti\ka}}}{6}+ {\ti F}_{3}(-{\ti\ka}^{-1})
\right\rbrack & \mbox{for ${\ti\ka} >1$},
\end{array}
\right. \label{56}
\eqn
and the function ${\ti F}_{s}$ is the {\it Riemann-Fermi-Dirac} 
type, such that
\beq
{\ti F}_{s}({\ti z})=\sum_{n=1}^{\infty} \frac{{\ti z}^n}{n^s }.\label{57}
\eeq 
Since we have a term ${\ti\om}_+{\ti\om}_-$ in the denominator
of eq.(60), we note that 
\beq
{\ti\om}_+{\ti\om}_-=\om_0^2(2+m\om_c\te)^2.\lb{pro}
\eeq 
We observe that eq.(60) shows a singularity at a critical point. So
we are now forced to distinguish two different cases. The first one, 
$m\om_c\te\neq -2$, is equivalent actually to ${eB\te\ov c}\neq 2$. 
Then we have $m\om_c\te>$ or $<-2$,
since there is a square, we can only discuss the global case.
Second one is a critical point $m\om_c\te=-2$
where eq.(60) diverges. Remembering that
by using eq.(\ref{pj}), we find that our cases
coincide with those noted in \cite{pol2}, namely
$B\te\neq 1$ and $B\te=1$. In this subsection, we assume 
that the former case holds in the further analysis.
However, the latter case will deal with in
the last subsection.

Let us examine eq.(60) in different limits of temperature
and by putting the condition: $m\om_c\te\neq -2$. In other words,
we want to derive
the thermodynamical potential and the related 
physics quantities at high and low
temperature at noncommutative level and 
compare with the standard case.\\ 

\no{\it i-\underline{High temperature limit}}\\
In this case we make the assumption $\vert \mu \pm \hbar {\ti\om}/2 \vert
\gg \be$  and we get
${\ti\ka}_{\pm} \approx 1$. Therefore using eqs.(59) and (60), $\Om_{\te}$
can be approximated by
\beq 
\Om_{\te} \approx \frac{4}{\be^3\hbar^2}\; {F_3(-1)\ov {\ti\om}_+{\ti\om}_-},\label{64}
\eeq
where $F_3(-1)=-0.901543$. In terms of $\te$ we have
\beq
\Om_{\te}\approx -0.901543\times 4\,{\left(\frac{1}{\be\hbar\om_0}\right)^2
\ov \be(2+m\om_c\te)^2}.\label{65}
\eeq
We remark from the last formula that $\pa_{\mu}\Om_{\te}=0$, 
namely there is no exchange
of electrons. This means that at high temperature, the
present system can be described as a canonical ensemble. However, 
the magnetization and susceptibility can be evaluated in this
case. We get for $M_{\te}$  
\beq
\bra{l}
M_{\te}= 0.901543\times 8\; \left(\frac{1}{\be\hbar\om_0}\; \right)^2
\;{e\te/\be c\ov(2+m\om_c\te)^3 },\lb{66}
\era
\eeq
and remembering the relation $\chi_{\te}={\pa M_{\te}\ov\pa B}$, 
we obtain for the susceptibility 
\beq
\bra{l}
\chi_{\te}=-0.901543\times
24\; \;{1\ov\be^3}\;\Big({e\te\ov\hbar c\om_0}\Big)^2\;
{1\ov (2+m\om_c\te)^4 }.
\era
\eeq
Let us examine some particular cases of the last 
equation. For a zero magnetic field, we find
\beq
\bra{l}
\chi_{\te}=-0.901543\times {3\ov 2}\; 
{1\ov\be^3}\;\Big({e\te\ov\hbar c\om_0}\Big)^2.
\lb{68}
\era
\eeq
This latter shows that $\chi_{\te}$ is $\te$-dependent. Therefore,
we have {\it Landau} diamagnetism since $\te$ is a real value. 
However, when $\te$ vanishes
there is no magnetic behaviour. This means that 
\beq
\chi_{\te=0}=0,\lb{69}
\eeq
which is compatible with the standard case. 
It is interesting to note that at high
temperature the system presents a magnetic 
behaviour in terms of $\te$ and it is
the canonical one. This effect does not 
appear in the commutative case. 
This is one of the original results derived in this paper.\\

\no{\it ii-\underline{Low temperature limit}}\\
Let us consider another interesting case, namely
$\mu \gg \hbar {\ti\om}/2$ and $\mu \ll \be$. 
In this situation, $\phi({\ti\ka})$ can be expressed as 
\beq
{\phi}({\ti\ka}_{\pm}) = {\ti A} \mp \frac{\ti\Delta}{2} + 
{\ti S}_{\pm}, \lb{70}
\eeq
and we have
\beq
\bra{l} 
{\ti A}= -2\mu\;{\frac{1}{3}
\left(\frac{\mu}{\hbar\om_0}\right)^2 + 
\frac{1}{4}\left(\frac{{\ti\om}}{\om_0}\right)^2+\frac{\pi^2}{3}
\left(\frac{1}{\be\hb\om_0}\right)^2\ov (2+m\om_c\te)^2},\\ 
\frac{\ti\Delta}{2} = 
2\hbar{\ti\om}{\frac{1}{2}
\left(\frac{\mu}{\hbar\om_0}\right)^2+ \frac{1}{24}\left(\frac{{\ti\om}}{\om_0}\right)^2+\frac{\pi^2}{6}
\left(\frac{1}{\be\hbar\om_0}\right)^2\ov (2+m\om_c\te)^2}, \\ 
{\ti S}_{\pm} =\left(\frac{1}{\be\hbar\om_0}\right)^2\;
\frac{4}{\be(2+m\om_c\te)^2}\; 
F_3(-\exp{[-\be( \mu \pm \hbar{\ti\om} /2)]}). \lb{71}
\era
\eeq
At low temperature,  ${\ti S}_{\pm}$ can be approximated by the following
relation 
\beq
{\ti S}_{0}= \left(\frac{1}{\be\hbar\om_0}\right)^2\;F_3(-e^{-\be\mu})\;
\frac{4}{\be(2+m\om_c\te)^2},
\eeq
and we can see immediately that this equation
\beq
\frac{\ti\Delta}{\vert {\ti A} + {\ti S}_{0} \vert} = 
\frac{\hbar {\ti\om}}{\mu} \left\lbrack \frac{ 3 + 
\pi^2 \left(\frac{1}{\be\mu}\right)^2 + \frac{1}{4} 
\left(\frac{\hbar{\ti\om}}{\mu}\right)^2}{ 1 + \pi^2 
\left(\frac{1}{\be\mu}\right)^2 + \frac{3}{4} 
\left(\frac{\hbar {\ti\om}}{\mu }\right)^2 - \left(\frac{1}
{\be\mu}\right)^3 F_3( -e^{-\be \mu}) } \right\rbrack\lb{73}
\eeq
tends to zero. Therefore, the thermodynamical potential can be written as follows
\beq
\bra{l}
\Om_{\te} =-\frac{2\mu}{(2+m\om_c\te)^2}\left\lbrack\frac{1}{3}\left(\frac{\mu}{\hbar\om_0}\right)^2+
{1\ov 4}\left(\frac{{\ti\om}}{\om_0}\right)^2 +\frac{\pi^2}{3}\left(\frac{1}{\be\hbar\om_0}\right)^2 
 -{2\ov \be\mu}\left(\frac{1}{\be\hbar\om_0}\right)^2F_3(-e^{-\be\mu})\right\rbrack. \lb{74}
\era
\eeq
In this quasiclassical regime, the average number of electrons is 
\beq
\langle {\ti N}_e \, \rangle = 4{\left({\mu/\hb\om_0}\ov 2+m\om_c\te\right)^2}
 \left\lbrack \frac{1}{2}+\frac{1}{8}\left(\frac{\hbar{\ti\om}}{\mu}\right)^2 
+\frac{\pi^2}{6}\left(\frac{1}{\be\mu}\right)^2
+\left(\frac{\om_0}{{\ti\om}}\right)^2(2+m\om_c\te)^2 \left(\frac{1}
{\be\mu}\right)^2 F_2 (-e^{-\mu \be}) \right\rbrack,
\eeq
which can be estimated as 
\beq
\langle {\ti N}_e \, \rangle \approx  2\left({\mu\ov\hb\om_0}\right)^2{1\ov (2+m\om_c\te)^2}.
\eeq
Notice that the  average number of electrons at low $T$ 
in the commutative case can be recovered 
just by switching off one of the parameters $B$ or $\te$. By using the
definition of 
magnetic moment, we obtain
\beq
\bra{l}
M_{\te} = -4{e\mu\te/c\ov (2+m\om_c\te)^3}\left\lbrack\frac{1}{3}\left(\frac{\mu}{\hbar\om_0}\right)^2+
{1\ov 4}\left(\frac{{\ti\om}}{\om_0}\right)^2 +\frac{\pi^2}{3}\left(\frac{1}{\be\hbar\om_0}\right)^2 
 -{2\ov\be \mu}\left(\frac{1}{\be\hbar\om_0}\right)^2F_3(-e^{-\be\mu})\right\rbrack\\
\qquad\;\;\;\;+{e\mu/mc\ov (2+m\om_c\te)^2}\left\lbrack {\om_c\ov \om_0^2}+
{m\te\ov 4\om_0^2}(2\om_c^2+\om^2)
+2{\om_c\ov\om_0^2}\left({m\om\te\ov 4}\right)^2\right\rbrack.
\era
\eeq
Therefore, the susceptibility takes the following form
\beq
\bra{l}
\chi_{\te} = 12\mu{(e\te/c)^2\ov (2+m\om_c\te)^4}\left\lbrack\frac{1}{3}
\left(\frac{\mu}{\hbar\om_0}\right)^2+{1\ov 4}\left(\frac{{\ti\om}}{\om_0}\right)^2 
+\frac{\pi^2}{3}\left(\frac{1}{\be\hbar\om_0}\right)^2 
-{2\ov\be\mu}\left(\frac{1}{\be\hbar\om_0}\right)^2F_3(-e^{-\be\mu})\right\rbrack\\
\qquad\;\;-4\mu\left({e\ov mc}\right)^2{m\te\ov (2+m\om_c\te)^3}
\left\lbrack {\om_c\ov \om_0^2}+{m\te\ov 4\om_0^2}(2\om_c^2+\om^2)
+2{\om_c\ov\om_0^2}\left({m\om\te\ov 4}\right)^2\right\rbrack\\
\qquad\;\;+\left({e\ov mc\om_0}\right)^2{\mu\ov (2+m\om_c\te)^2}
\left\lbrack 1+2m\om_c\te+6\left({m\om\te\ov 4}\right)^2\right\rbrack.
\era
\eeq 
For zero magnetic field, we find
\beq
\chi_{\te}= \chi_p\left\lbrack 1+(m\om_0\te)^2\Bigg\{1+
\left(\frac{\mu}{\hbar\om_0}\right)^2+3\left(\frac{{m\om_0\te}}{2}\right)^2 
+\left(\frac{\pi}{\be\hbar\om_0}\right)^2 
-{6\ov\be \mu}\left(\frac{1}{\be\hbar\om_0}\right)^2F_3(-e^{-\be\mu})\Bigg\}\right\rbrack,
\eeq
which implies that a correction is obtained in this case. Let us
solve the above 
equation in order to obtain the limiting cases for $\chi_{\te}$. So,
Eq.(78)
can be written in compact form 
 \beq
\chi_{\te}= \chi_p\left\lbrack 1+a\la+{3\ov 4}\la^2\right\rbrack,
\eeq
where $\la=(m\om_0\te)^2$ and $a=1+
\left(\frac{\mu}{\hbar\om_0}\right)^2+3\left(\frac{{m\om_0\te}}{2}\right)^2 
+\left(\frac{\pi}{\be\hbar\om_0}\right)^2 
-{1\ov\be \mu}\left(\frac{1}{\be\hbar\om_0}\right)^2F_3(-e^{-\be\mu})$. 
The possible solutions of eq.(79) are 
\beq
\la_{\pm}={2\ov 3}(-a\pm\sqrt{a^2-3}),
\eeq
we can see that at $\la_{\pm}$ values, the susceptibility vanishes. However for $\la\in
]\la_-,\la_+[$, there is a diamagnetic behaviour, but otherwise the system exibits a
paramagnetic behaviour. Now by switching off the noncommutative parameter,
we get
\beq
\chi_{\te}\ev\chi_p = \mu \left(\frac{e}{2mc\om_0}\right)^2,
\eeq
this shows that in the commutative case, the system exibits an orbital 
paramagnetism in the limiting case for magnetic field \cite{jel2}. 

\subsection{Fermi-Dirac trace formulas }
It is well known that, like the {\it Gaussian} function, 
the function $\mbox{sech}{x} = 1/\cosh{x}$ is a fixed point of 
the {\it Fourier} transform in 
the {\it Schwartz} space:
\beq
\frac{1}{\cosh{\sqrt{\frac{\pi}{2}}x}}
= \frac{1}{\sqrt{2\pi}} \int_{-\infty}^{+\infty} 
\frac{e^{-ixy}}{\cosh{\sqrt{\frac{\pi}{2}}y}}\, dy.
\eeq
Then for a given Hamiltonian ${H}$, the {\it Fermi} operator is
\beq
f({H})\equiv\frac{1}{1+e^{\be({H}-\mu)}}=
\int_{-\infty}^{+\infty}\frac{e^{-(ik+1) \frac{\be}{2} ({H}-\mu)}}
{4\cosh{\frac{\pi}{2}k}}\, dk, \label{a.2}
\eeq
and the corresponding thermodynamical potential operator takes the form
\beq
-\frac{1}{\be}\log{(1+e^{-\be({H} - \mu)})} = -\frac{1}{\be}\int_{-\infty}^{+\infty} \frac{e^{-(ik +
1)\frac{\be}{2}({H} - \mu)}}{(2\cosh{\frac{\pi}{2}k})(ik+1)}\, dk.
\eeq
Therefore, the average number of fermions and the thermodynamical potential  
can be written as follows 
\beq
\langle N \rangle = \mbox{Tr}f({H})  
=\int_{-\infty}^{+\infty} 
\frac{e^{(ik+1)\frac{\be\mu}{2}}}{4\cosh{\frac{\pi}{2}k}}\Theta(k)\,dk, 
\label{a.4} 
\eeq
\beq 
\Om = \mbox{Tr}(-\frac{1}{\be}\log{(1+ e^{-\be({H} - \mu)})})  
=-\frac{1}{\be}\int_{-\infty}^{+\infty} 
\frac{e^{(ik+1)\frac{\be\mu}{2}}}{(2\cosh{\frac{\pi}{2}k})(ik+1)}\Theta(k)\,dk, 
\label{a.5} 
\eeq 
where $\Theta$ designates the function  
\beq
\Theta (k) =\mbox{Tr} (e^{-(ik +
1)\frac{\be}{2}{H}}).
\label{a.6}
\eeq
Observe that  $(2n+1)i, \ n\in \Z$ are (simple) poles for the function 
$1/\cosh{\frac{\pi}{2}k}$ and $i$ is a pole for the functions $\Theta(k)$ 
and $1/(ik +1)$.
These {\it Fourier} integrals can be 
evaluated by using residue theorems  
if the integrand functions 
$\Phi_1(k)=\Theta (k)/\cosh{\frac{\pi}{2}k}$ and  
$\Phi_2(k)=\Theta (k)/((ik+1)\cosh{\frac{\pi}{2}k})$
satisfy the {\it Jordan} Lemma, that is,   
$\Phi_1(r e^{i\vr}) \leq g(r)$, $\Phi_2(r e^{i\vr})\leq h(r)$, for all 
$\vr \in \lbrack 0, \pi \rbrack$, and 
$g(r)$ and $h(r)$ vanish as $r\to \infty$. The quantities
$\langle N \rangle$ and $\Om$ are then formally given by
\beq
2\pi i \left\lbrack a_{-1}(i)+\sum_{n=1}^{\infty}a_{-1}((2n+1)i) 
+\sum_{\nu} a_{-1}(k_{\nu})  \right\rbrack, 
\label{a.7}
\eeq
where $a_{-1}(\cdot)$ denotes the residue of the involved integrand at pole 
$(\cdot)$, and the $k_{\nu}$'s are the poles (with the exclusion of the pole $i$) of $\Theta(k)$ in the
complex $k$-plane.\\

We now apply the above tools to get the thermodynamical potential through
{\it Fermi-Dirac} trace formulas. To do that, we begin by evaluating
eq.(87) at noncommutative level. Then, in our case we can write
$\Theta(k)$ as follows
\beq
{\ti\Theta}(k)= e^{-(ik+1)\frac{\be}{4}\hbar{\ti\om}}\, 
\frac{1}{1 - e^{-(ik + 1)\frac{\be}{2}\hbar {\ti\om}_+}}\, 
\frac{1}{1 - e^{-(ik + 1)\frac{\be}{2}\hbar {\ti\om}_-}}.
\eeq
Subsequently, the {\it Fourier} integral representation for the 
thermodynamical 
potential eq.(86) becomes
\beq
\Om_{\te} = -\frac{1}{\be }\int_{-\infty}^{+\infty} 
\frac{e^{-(ik+ 1)\frac{\be}{2}(\frac{\hbar{\ti\om}}{2}-\mu)}}
     {2\cosh{\frac{\pi}{2}k}} \, 
\left(\frac{1}{ik+1}\right)
\left(\frac{1}{1-e^{-(ik+ 1)\frac{\be}{2}\hbar {\ti\om}_+}}\right) 
\left(\frac{1}{1 - e^{-(ik + 1)\frac{\be}{2}\hbar {\ti\om}_-}}\right) dk.
\eeq
As indicated in the formula (\ref{a.6}), this {\it Fourier} 
integral is 
given 
as a series  by using the residue theorem. One can easily see  that the 
numbers $(2n+1)i$, $n\in \Z$ are simple poles of 
$\mbox{sech}{\frac{\pi}{2}k}$, $i$~is a double pole of $\Theta (k)$,
and $i+4\pi n/(\be\hbar{\ti\om}_+)$, $i+4\pi n/(\be\hbar{\ti\om}_-)$, 
$n\in \Z^{\ast}$ are simple or double poles of $\Theta(k)$ according to
whether ${\ti\om}_+$ and ${\ti\om}_-$ are uncommensurable or not. 
In order to fulfill the requirements of the {\it Jordan} Lemma, 
one has to consider the following two cases: 
$\mu\leq\hbar{\ti\om}/2$ and $\mu\geq\hbar{\ti\om}/2$. 
In the first case we take an integration path lying in the lower half-plane and
involving only the simple poles $(2n+1)i$, $n<0$. 
We get 
\beq
\Om_{\te} = \frac{1}{4\be}\sum_{n=1}^{\infty} \frac{(-1)^n}{n}
\frac{e^{\beta \mu n}}
     {\sinh{(\frac{\be}{2}\hbar {\ti\om}_+n)}\,
\sinh{(\frac{\be}{2}\hbar {\ti\om}_-n)}}.
\label{4.20}
\eeq 
In the second case, an integration path in the upper half-plane is chosen. 
It encircles all the other poles: $(2n+1)i$, $n\geq 0$, 
$i+4\pi n/(\be\hbar{\ti\om}_+)$, $i+4\pi n/(\be\hbar{\ti\om}_-)$, $n\in\Z^{\ast}$.
We present the result in a manner which will render apparent the various regimes:
\beq
\begin{array}{cccccccc}
\Om_{\te} &=& &({\ti\Om}_L + {\ti\Om}_{01})& + &{\ti\Om}_{02}& + 
&{\ti\Om}_{\mbox{\scriptsize osc}}\\
&=&2\pi i(&\overbrace{a_{-1}(i)}& +&\overbrace{\sum_{n\geq 1}
a_{-1}((2n+1)i)}& 
 +&\overbrace{\sum_{n_{\pm}\not= 0}(a_{-1}(i+\frac{4\pi}
{\be\hbar{\ti\om}_{\pm}}n_{\pm})}).
\end{array}
\eeq
Here we suppose that
$m\om_c\te\neq -2$ is satisfied and as mentioned before
the opposite case will be considered in the last 
subsection.
For ${\ti\Om}_L$, we find
\beq
{\ti\Om}_L=\frac{\mu}{6\om_0^2}\left(\frac{\om_c+(m\om_c^2\te/4)-
m\om_0^2\te}{2+m\om_c\te}\right)^2,
\eeq
and ${\ti\Om}_{01}$ can be written as follows
\beq
{\ti\Om}_{01}=-\frac{2\mu}{(2+m\om_c\te)^2}\left\lbrack\left(\frac{\mu}
{\hbar \om_0} \right)^2 
 +\left(\frac{\pi}{\be\hbar\om_0}\right)^2\right\rbrack+\frac{\mu}{12}.
\eeq 
${\ti\Om}_{02}$ is given by
\beq
{\ti\Om}_{02}=\frac{1}{4\be}\sum_{n=1}^{\infty}\frac{(-1)^n}{n}
\frac{\exp{(-\be\mu n)}}
{\sinh{(\frac{\be\hbar {\ti\om}_+}{2}n)}\,\sinh{(\frac{\be\hbar 
{\ti\om}_-}{2}n)}}.
\eeq
For ${\ti\om}_+/{\ti\om}_- \not\in \Q$, we obtain 
\beq
\bra{l} 
{\ti\Om}_{\mbox{\scriptsize osc}}= 
\frac{1}{2\be} \sum_{n=1}^{\infty} \frac{(-1)^n}{n} 
\left\lbrack\frac{\sin{(\frac{2 \mu}{\hbar {\ti\om}_-}\pi n)}}
{\sin{(\frac{{\ti\om}_+}{\om_-}\pi n)}\,
\sinh{(\frac{2\pi^2 n}{\be\hbar{\ti\om}_-})}} 
+ \frac{\sin{(\frac{ 2\mu}{\hbar {\ti\om}_+}\pi n)}}
{\sin{(\frac{{\ti\om}_-}{{\ti\om}_+}\pi n)}\,
\sinh{(\frac{2 \pi^2 n}{\hbar {\ti\om}_+})}} \right\rbrack \\ 
\qquad\; \equiv {\ti\Om}_{\mbox{\scriptsize osc}}^- + 
{\ti\Om}_{\mbox{\scriptsize osc}}^+.
\era
\eeq
However for  ${\ti\om}_+/{\ti\om}_-=p/q\in\Q,\ \gcd{(p,q)} = 1, \ 
{\ti\om}_+/p = {\ti\om}_-/q = 2l/(\hbar\be) \in\RR$, we have 
\beq
\bra{l}
{\ti\Om}_{\mbox{\scriptsize osc}} =  \frac{1}{2\be} \left\lbrack 
\sum_{n=1 ,\, n\not\equiv 0\, \mbox{\scriptsize mod}\, q}^{\infty} 
\frac{ (-1)^n}{n} \frac{\sin{(\frac{ 2\mu}{\hbar {\ti\om}_-}\pi n)}}
{\sin{(\frac{{\ti\om}_+}{\om_-}\pi n)}\,
\sinh{(\frac{2\pi^2 n}{\be\hbar {\ti\om}_-})}} 
\right.\\
 \qquad\;\;\;\; +  \sum_{n=1 ,\, m\not\equiv 0 \, 
\mbox{\scriptsize mod}\, p}^{\infty} 
\frac{ (-1)^n}{n}\frac{\sin{(\frac{ 2\mu}{\hbar {\ti\om}_+}\pi n)}}
{\sin{(\frac{{\ti\om}_-}{{\ti\om}_+}\pi n)}\,
\sinh{(\frac{2\pi^2 n}{\hbar {\ti\om}_+}\pi^2 n)}} 
\label{4.26}
\\
\qquad\;\;\;\; + \left. \frac{1}{lpq}\sum_{k=1}^{\infty} 
\frac{(-1)^{(p+q)k}}{k\sinh{(\frac{\pi^2}{l}k)}}\left\lbrack
\be\mu\cos{(\frac{\be\mu\pi k}{l})} 
-(\pi\coth{(\frac{\pi^2}{l}k)} + \frac{l}{\pi k})
\sin{(\frac{\be\mu\pi k}{l})} \right\rbrack \right\rbrack . 
\era
\eeq 

Therefore, the average number of electrons is 
\bqn
\nonumber \langle {\ti N}_e\,\rangle &=& 
-\frac{1}{6\om_0^2}\left(\frac{\om_c+(m\om_c^2\te/4)-
m\om_0^2\te}{2+m\om_c\te}\right)^2  
+\frac{1}{2}\left\lbrack{4\ov (2+m\om_c\te)^2}
\Bigg(\left(\frac{\mu}{\hbar \om_0}\right)^2 
+\frac{1}{3}\left(\frac{\pi}{\hbar\om_0}\right)^2\Bigg)
-\frac{1}{6}\right\rbrack \\
\nonumber & & +\frac{1}{4} \sum_{n=1}^{\infty} (-1)^n
\frac{e^{-\be \mu n}}{\sinh{(\frac{\be}{2}\hbar{\ti\om}_+n)}\,
\sinh{(\frac{\be}{2}\hbar {\ti\om}_-n)}} \\
\nonumber & &- \pi\sum_{n=1}^{\infty} (-1)^n 
\left\lbrack \frac{1}{\be\hbar{\ti\om}_-}
\frac{\cos{(\frac{2 \mu}{\hbar {\ti\om}_-}\pi n)}}
{\sin{(\frac{{\ti\om}_+}{{\ti\om}_-}\pi n)}\,
\sinh{(\frac{2\pi^2 n}{\be\hbar{\ti\om}_-})}} 
+\frac{1}{\be\hbar {\ti\om}_+ }\frac{\cos{(\frac{2\mu}{\hbar 
{\ti\om}_+}\pi n)}}
{\sin{(\frac{{\ti\om}_-}{{\ti\om}_+}\pi n)}\,
\sinh{(\frac{2\pi^2 n}{\hbar {\ti\om}_+})}} \right\rbrack \\
&\equiv& \langle {\ti N}_e\, \rangle_{L} + \langle 
{\ti N}_e\,\rangle_{01} 
+\langle {\ti N}_e\, \rangle_{02} + \langle {\ti N}_e\,
\rangle_{\mbox{\scriptsize osc}}^- 
+\langle {\ti N}_e\, \rangle_{\mbox{\scriptsize osc}}^+,\label{5.1}
\eqn
and the magnetic moment can be written as follows
\beq
 M_{\te} = {\ti M}_L +{\ti M}_{01}+{\ti M}_{02}+
{\ti M}_{\mbox{\scriptsize osc}}^- +
 {\ti M}_{\mbox{\scriptsize osc}}^+,
\eeq
where
\beq
\bra{l}
{\ti M}_L = -\frac{e\mu}{3mc\om_0^2}\left(\frac{\om_c+(m\om_c^2\te/4)-
m\om_0^2\te}{2+m\om_c\te}\right)\left\lbrack {1\ov 2}-
{(\om_c+(m\om_c^2\te/4)-m\om_0^2\te)m\te\ov (2+m\om_c\te)^2}
\right\rbrack,\\
{\ti M}_{01}=-\frac{4e\mu}{3mc}{m\te\ov (2+m\om_c\te)^3}
\left\lbrack\left(\mu\ov\hb\om_0\right)^2+
\left(\pi\ov\be\hb\om_0\right)^2\right\rbrack,
\era
\eeq
and for ${\ti M}_{02}$, we have
\beq
\bra{l}
{\ti M}_{02} = \frac{\hb e}{4mc}\sum_{n=1}^{\infty}
(-1)^n {e^{-\beta \mu n }\ov\sinh{(n\beta\hbar{\ti\om}_+)}
\sinh{(n\beta\hbar{\ti\om}_-)}}\\
\qquad\;\;\;\;\times \Bigg[
{1\ov{\ti\om}}\left(\om_c+{m\te\ov 4}(2\om_c^2+\om^2)+
2\om_c(m\om\te/4)^2\right)
\times\left(\coth{(n\beta\hbar{\ti\om}_+)}
+\coth{(n\beta\hbar{\ti\om}_-)}\right)\\
\qquad\;\;\;\; +{1\ov 2}(2+m\om_c\te)
\times\left(\coth{(n\beta\hbar{\ti\om}_+)}
-\coth{(n\beta\hbar{\ti\om}_-)}\right) \Bigg],
\era
\eeq
and, for the irrational case $\om_+/\om_- \not\in \Q$, 
\beq
\bra{l}
{\ti M}_{\mbox{\scriptsize osc}}^- = \frac{e\pi}{\be mc} 
\sum_{n=1}^{\infty}
 {(-1)^n\ov \sin(\pi n{{\ti\om}_+\ov {\ti\om}_-}) 
\sinh({2\pi^2n\ov\be\hbar{\ti\om}_-})}\Bigg[
{1\ov{\ti\om}}\left(\om_c+{m\te\ov 4}(2\om_c^2+\om^2)+
2\om_c(m\om\te/4)^2\right)\\
\qquad\;\;\;\;\times\left({\mu\ov\hb{\ti\om}_-^2}
\cos{(2\pi n{\mu\ov\hbar{\ti\om}_-})}
-{{\ti\om}_c\ov{\ti\om}_-^2}\cot(\pi n{{\ti\om}_+\ov {\ti\om}_-})         
\sin{(2\pi n{\mu\ov\hbar{\ti\om}_-})}-{\pi\ov\be\hb{\ti\om}_-^2}
\sin(\pi n{{\ti\om}_+\ov {\ti\om}_-}) \coth({2\pi^2n\ov\be\hbar{\ti\om}_-})
\right)\\
\qquad\;\;\;\;+{1\ov 2}(2+m\om_c\te)\times\Bigg(-{\mu\ov\hb{\ti\om}_-^2}
\cos{(2\pi n{\mu\ov\hbar{\ti\om}_-})}
+{{\ti\om}\ov{\ti\om}_-^2}\cot(\pi n{{\ti\om}_+\ov {\ti\om}_-})         
\sin{(2\pi n{\mu\ov\hbar{\ti\om}_-})}
\\ \qquad\;\;\;\; +{\pi\ov\be\hb{\ti\om}_-^2}
\sin(\pi n{{\ti\om}_+\ov {\ti\om}_-}) \coth({2\pi^2n\ov\be\hbar{\ti\om}_-})
\Bigg)\Bigg],
\era
\eeq
and the same result can be obtained for 
${\ti M}_{\mbox{\scriptsize osc}}^+$
\beq
\bra{l}
{\ti M}_{\mbox{\scriptsize osc}}^+ = \frac{e\pi}{\be mc} \sum_{n=1}^{\infty}
 {(-1)^n\ov \sin(\pi n{{\ti\om}_+\ov {\ti\om}_-}) 
\sinh({2\pi^2n\ov\be\hbar{\ti\om}_-})}\Bigg[
{1\ov{\ti\om}}\left(\om_c+{m\te\ov 4}(2\om_c^2+\om^2)+
2\om_c(m\om\te/4)^2\right)\\
\qquad\;\;\;\; \times\left({\mu\ov\hb{\ti\om}_+^2}
\cos{(2\pi n{\mu\ov\hbar{\ti\om}_+})}
-{{\ti\om}_c\ov{\ti\om}_+^2}\cot(\pi n{{\ti\om}_-\ov {\ti\om}_+})         
\sin{(2\pi n{\mu\ov\hbar{\ti\om}_+})}-{\pi\ov\be\hb{\ti\om}_+^2}
\sin(\pi n{{\ti\om}_-\ov {\ti\om}_+}) \coth({2\pi^2n\ov\be\hbar{\ti\om}_+})
\right)\\
\qquad\;\;\;\; 
+{1\ov 2}(2+m\om_c\te)\times\Bigg(-{\mu\ov\hb{\ti\om}_+^2}\cos{(2\pi 
n{\mu\ov\hbar{\ti\om}_+})}
+{{\ti\om}\ov{\ti\om}_+^2}\cot(\pi n{{\ti\om}_-\ov {\ti\om}_+})         
\sin{(2\pi n{\mu\ov\hbar{\ti\om}_+})}\\
\qquad\;\;\;\; +{\pi\ov\be\hb{\ti\om}_+^2}
\sin(\pi n{{\ti\om}_-\ov {\ti\om}_+}) \coth({2\pi^2n\ov\be\hbar{\ti\om}_+})
\Bigg)\Bigg].
\era
\eeq

Similar formulas can be derived for ${\cal M}^{\pm}_{\mbox{\scriptsize osc}}$
in the rational case. These expression can be studied in different limits 
of temperature, magnetic field and noncommutative parameter in order to 
understand the behaviour of the system under
consideration. This will be the subject of the forthcoming work \cite{jel3}.

\subsection{Critical point $m\om_c\te=-2$}
Let us mention that this critical point is actually
equivalent to ${eB\te\ov c}=-2$. By
using the transformation (\ref{pj}) and taking
$(c=1,e=1,m=1)$, we find the critical point
${B\te}=1$ obtained in \cite{pol2}.

By taking $m\om_c\te=-2$, the set of frequencies
defined in subsection $3.1$ becomes 
\beq
\bra{l}
{\ti\om}=-{\om^2\ov 2\om_c},\qquad {\ti\om}_c=
{\om_c\ov 2}(1+{4\om_0^2\ov \om_c^2}),\\
{\ti\om}_+=0,\qquad {\ti\om}_-=2{\ti\om}. 
\era
\eeq
Now if we come back to eq.(21), we get a Hamiltonian
of a harmonic oscillator of frequency ${\ti\om}$, such that
\beq
H_{\te}(\te=-{2\ov m\om_c})={\hb{\ti\om}\ov 2}(2{\ti N}_g+1),
\eeq
where the eigenstates and the eigenvalues
are $|{\ti n}_g>$ and ${\hb{\ti\om}\ov 2}(2{\ti n}_g+1)$, 
respectively. 
Therefore, the thermodynamical 
potential eq.(44) can
now be written in terms of 
$H_{\te}(\te=-{2\ov m\om_c})$ 
\beq
\Om_{\te}(\te=-{2\ov m\om_c})= -{1\ov \be}{\textbf{Tr}}   
e^{-\be (H_{\te}(\te=-{2\ov m\om_c})-\mu)},\lb{the1}
\eeq
where the corresponding partition function is
\beq
Z_{\te}(\te=-{2\ov m\om_c})= {\textbf{Tr}}\ 
e^{-\be (H_{\te}(\te=-{2\ov m\om_c})-\mu)},\lb{par1}
\eeq
and the trace is taken on the eigenstates
$|{\ti n}_g>$. 
Actually, we can construct coherent states in such a way that
\beq
\bra{l}
\mid {\ti z}_{g}\rangle = \exp{\lbrack-\frac{1}{2}
\vert {\ti z}_{g}\vert^2\rbrack }\,
 e^{{\ti z}_{g} {\ti a}_{g}^{\dagger}}
\mid {\ti 0}\rangle, \\ \lb{ngcs}
{\ti a}_{g} \mid {\ti z}_{g} \rangle = 
{\ti z}_{g}\mid {\ti z}_{g} \rangle.
\era
\eeq
With the respect of the last equation,  
$Z_{\te}(\te=-{2\ov m\om_c})$ can be expressed  as follows
\beq
Z_{\te}(\te=-{2\ov m\om_c})=e^{\be\mu} \int d^2 {\ti z}_{g}
<{\ti z}_{g}|
e^{-\be {\hb{\ti\om}\ov 2}(2{\ti N}_g+1)}|{\ti z}_{g}>.\lb{par2}
\eeq
To calculate the partition function,
one can consider the boson-operator identity \cite{lou}
\beq
e^{\xi {a}^{\da}{a}}=
\sum_{n=0}^{\infty}{(e^{\xi}-1)^n\over n!} 
{a}^{\da}{ a},
\eeq
which holds for any operators ${a}^{\da}$ and $a$
satisfying the commutation relation $[{a},a^{\da}]=1$. By
applying this identity, we can show that 
\beq
Z_{\te}(\te=-{2\ov m\om_c})= e^{-\be ({\hb{\ti\om}\ov 2}-\mu)}
\int d^2 {\ti z}_{g}
e^{-|{\ti z}_{g}|^2(1-e^{-\be\hb{\ti\om}})}.\lb{par3}
\eeq
After integration, we obtain
\beq
Z_{\te}(\te=-{2\ov m\om_c})= 
{e^{\be\mu}\over 4\sinh\Big({\be\hbar{\ti\om}\over 2}\Big)}.\lb{par4}
\eeq
Thus, the thermodynamical potential becomes
\beq
\Om_{\te}(\te=-{2\ov m\om_c})=-{1\ov\be}
\log{\Big(4\sinh({\be\hbar{\ti\om}\over 2})\Big)}
-\mu.\lb{pot}
\eeq
We get for the magnetic moment
\beq
M_{\te}(\te=-{2\ov m\om_c})=-{e\hb\ov 4mc}{\om^2\ov\om_c^2}
\coth{({\be\hbar\om^2\over 4\om_c})},\lb{mag}  
\eeq
and hence the susceptibility is 
\beq
\chi_{\te}(\te=-{2\ov m\om_c})={1\ov 2}\Big({e\hb\ov 2mc}\Big)^2
\Big[{1\ov \hb\om_c}({\om^2\ov\om_c^2}-1) 
\coth{({\be\hbar\om^2\over 4\om_c})}
+{\be\ov 2}{\om^4\ov\om_c^4}(1+
\coth^2{({\be\hbar\om^2\over 4\om_c})})\Big].\lb{sus}
\eeq
From the last equation, 
we observe that susceptibility 
becomes infinite at zero magnetic field. Note that, there
are some physical systems where infinite susceptibility
is actually seen \cite{ber}.
 
\section{Conclusion} 
We have investigated the {\it Fock-Darwin} Hamiltonian on the noncommutative 
space. We started by giving 
a noncommutative version of this Hamiltonian. Subsequently, the 
eigenstates
and the corresponding eigenvalues has been derived through two methods, 
an algebraic and an analytical. The degeneracy of 
{\it Landau} levels has
been considered and some algebras: $su(2)$ and $su(1,1)$ have been
realized. In particular it has been shown that the
degeneracy of {\it Landau} levels
can be lifted for this model at weak magnetic field limit. Using the {\it
Berezin-Lieb} inqualities,
we have obtained the magnetic behaviour of this model at high
temperature, which is abscent
in the commutative case. For low temperature, a 
$\te$-dependent correction to
susceptibility
has been pointed out. Furthermore, through the
use of the {\it Fermi-Dirac} 
trace formulas, a generalization of the thermodynamical potential,
the average number of electrons
and the magnetic moment has been found in terms of the noncommutative
parameter. At critical point, by using another approach, the 
magnetic moment and susceptibility have been obtained.

Finally, we mention that this generalization can be studied in various
regimes of temperature, magnetic field and noncommutative parameter. 
We could think also to investigate the relationship between the spatial
distribution
of current and the magnetic moment of 
the whole system at the noncommutative level. 
Another possibility is to study
the results derived in
this paper numerically.  
We hope to return to these questions in a subsequent publication.

\section*{Acknowledgements}
I would like to thank G. Maiella's group for the kind invitation
to visit INFN Naples Section, where the present work was finished.
This work was supported in part by High Energy Section at
AS-ICTP. I am grateful to \"O.F. Dayi and J.P. Gazeau with
whom we collaborated, separately, on the papers \cite{jel1}
and \cite{jel2}. I am thankful to L.  Boubekeur and B.A. Hammou 
for many discussions about noncommutativity. I am also grateful 
to L. Cappiello, C. Saclioglu and G.~Thompson for a careful reading of the 
manuscript and A. Polychronakos for his remarks. I~wish to 
thank my friend A. Elmhamdi for his kind 
hospitality during my stay at Trieste. I~am indebted to the referees 
for interesting comments and suggestions.

\end{document}